\documentclass[ajl]{emulateapj}

\usepackage{graphicx,epsfig,natbib,color}
\usepackage{lscape}
\usepackage{graphicx}
\usepackage{epsfig}
\usepackage{natbib}

\begin{document}
\title{Differential Emission Measure Analysis of A Limb Solar Flare on 2012 July 19}

\author{J. Q. Sun\altaffilmark{1,2}, X. Cheng\altaffilmark{1,2}, M. D. Ding\altaffilmark{1,2}}

\affil{$^1$ School of Astronomy and Space Science, Nanjing University, Nanjing 210093, China}\email{dmd@nju.edu.cn; xincheng@nju.edu.cn}
\affil{$^2$ Key Laboratory for Modern Astronomy and Astrophysics (Nanjing University), Ministry of Education, Nanjing 210093, China}

\begin{abstract}
We perform Differential Emission Measure (DEM) analysis of an M7.7 flare that occurred on 2012 July 19 and was well observed by the Atmospheric Imaging Assembly (AIA) aboard the \emph{Solar Dynamic Observatory}. Using the observational data with unprecedented high temporal and spatial resolution from six AIA coronal passbands, we calculate the DEM of the flare and derive the time series of maps of DEM-weighted temperature and emission measure (EM). It is found that, during the flare, the highest EM region is located in the flare loop top with a value varying between $\sim$ $8.4\times10^{28}$ $\mathrm{cm}^{-5}$ and $\sim$ $2.5\times10^{30}$ $\mathrm{cm}^{-5}$. The temperature there rises from $\sim$ 8 MK at about 04:40 UT (the initial rise phase) to a maximum value of $\sim$ 13 MK at about 05:20 UT (the hard X-ray peak). Moreover, we find a hot region that is above the flare loop top with a temperature even up to $\sim$16 MK. We also analyze the DEM properties of the reconnection site. The temperature and density there are not as high as that in the loop top and the flux rope, indicating that the main heating may not take place inside the reconnection site. In the end, we examine the dynamic behavior of the flare loops. Along the flare loop, both the temperature and the EM are the highest in the loop top and gradually decrease towards the footpoints. In the northern footpoint, an upward force appears with a biggest value in the impulsive phase, which we conjecture originates from chromospheric evaporation.   
\end{abstract}


\section{INTRODUCTION}
A solar flare is one of the most violent eruptive phenomena in the solar corona. It is generally accepted that the energy released during the flare is pre-stored in magnetic field and magnetic reconnection plays an essential role in converting magnetic energy into various energy forms like thermal energy of the plasma, kinetic energy of accelerated particles, and emissions in almost all wavelengths. In the past decades, a standard flare model (CSHKP; \citealt{Carmichael1964a, Sturrock1966a, Hirayama1974a, Kopp1976a}) has been established. It can explain many observational properties of flares such as two separating ribbons, formation of a cusp-shaped structure, etc. Even though the standard model has made a great achievement, the detailed process of flare energy release is still not clear, especially how and where the magnetic energy is most effectively converted to other forms of energy.\par

When a flare occurs, the flare atmosphere experiences different heating processes such as Joule heating (e.g. \citealt{Spicer1981b, Spicer1981a, Holman1985a}), shock heating (e.g. \citealt{Petschek1964a, Tsuneta1997a}), electron (e.g. \citealt{Fletcher1995a, Fletcher1996a, Fletcher1998a}) or proton (e.g. \citealt{Voitenko1995a, Voitenko1996a, Voitenko1999a}) beam heating, radiative backwarming (e.g. \citealt{Hudson1972a, Metcalf1990a, Metcalf1990b, Ding1996a}), and inductive current heating (e.g. \citealt{Melrose1995a, Melrose1997a}). These heating processes work with different efficiencies in different locations of the flare such as the reconnection site, flare loop top, and footpoints. Until now, to clarify the specific heating processes in flares is still an open question. Therefore, a quantitative assessment of the structure and evolution (particularly the temperature and density) of the flare region are critical to determine when, where and which heating processes are taking place.\par 

Previous studies on temperature and density of flares are mainly based on X-ray spectral observations (e.g. \citealt{Milkey1971a, Horan1971a, Dere1974a, Dere1977a, Cheng1977a, Landini1979a, Duijveman1983a, Denton1984a, Bornmann1985a}). \cite{Kahler1970a} fitted a thermal model using the X-ray data from OGO-5, and obtained the evolutions of temperature and emission measure (EM) in a flare. The temperatures they derived vary from about 5 MK to more than 10 MK and peak earlier than the EM. \cite{Dere1979a} used the line ratio method to derive the temperature and electron density, and found that when the temperature of flares rises from 1 to 10 MK, the electron density also rises from $1.0\times10^{10}$ to $5.0\times10^{11}$ $\mathrm{cm}^{-3}$. \cite{Doschek1981a} also used the line ratio method to obtain the electron temperatures of two M-class flares. Meanwhile, they derived the electron density using the O \uppercase\expandafter{\romannumeral7} lines. For both flares, they found the peak temperature and peak density are 18 MK and $10^{12}$ $\mathrm{cm}^{-3}$, respectively. Statistical researches were done by \cite{Feldman1996b, Feldman1996a}, who showed that the temperature and the volume emission measure range from 4 to 25 MK and from $10^{46}$ to $10^{50}$ $\mathrm{cm}^{-3}$, respectively, for a sample of more than 860 (A2 to X2 class) flares. \par

With the launch of Yohkoh, the multi-wavelength imaging observations make it possible to derive the two-dimensional temperature and EM maps of flares. For example, \cite{McTiernan1993a} obtained maps of temperature and EM of flares for the first time even though with a low spatial resolution. Results with higher resolution were then derived through the data from RHESSI (e.g. \citealt{Li2007a}), Hinode (e.g. \citealt{Reeves2009a, Winebarger2011a, Hahn2011a, Graham2013a}), and \emph{Solar Dynamics Observatory} (\emph{SDO}) (e.g. \citealt{Hannah2012a, Aschwanden2013a, Plowman2013a}). The time evolution of DEM in different regions of a flare was studied by \cite{Battaglia2012a} using the \emph{SDO}/Atmospheric Imaging Assembly (\emph{SDO}/AIA) data.\par
 
In this paper, we investigate the structure and evolution of the flare on 2012 July 19 using the \emph{SDO}/AIA data with unprecedented temporal and spatial resolution. We derive quantitatively the temperature and EM through the DEM analysis, which provide critical information to understand the energy and heating processes of flares. An overview of the flare is presented in Section 2. The data reduction and the DEM method are introduced in Section 3. The results are shown in Section 4, which is followed by a summary and conclusion in Section 5. \par

 \section{OBSERVATIONS}
 
On 2012 July 19, an M7.7 flare occurred at the solar west limb in NOAA active region 11520. Figure 1 shows the RHESSI hard X-ray (HXR) flux, GOES 1-8  {\AA} soft X-ray (SXR) flux, and AIA integrated EUV flux of the flare. As shown in Figure 1, the SXR flux started to increase at about 04:17 UT and peaked at about 05:58 UT. The gradual phase lasted more than 8 hours. At about 05:10 UT, a CME eruption was clearly visible. We divide the flare evolution into four phases: the pre-flare phase, the initial rise phase, the impulsive phase, and the gradual phase, as shown in Figure 1(b). \par

At the beginning of the flare ($\sim$ 04:17 UT), the whole loop structure first appeared in the 94 {\AA} and 131 {\AA} images, and a flux rope rose and expanded \citep{Patsourakos2013a}. While there were no loop structures visible in other low temperature channels; therefore, the temperature of the loop is about 7--10 MK according to the temperature response curves of these two filters \citep{ODwyer2010a}. In the initial rise phase, from 04:17 to 05:10 UT, the GOES SXR flux experienced an increase at a slow rate, indicating that the magnetic reconnection rate was relatively slow. As \cite{LiuWei2013a} concluded, the probable reconnection site is located between the flare loop top and the flux rope, which are both heated by the reconnection outflows. The continuous reconnection may lead to the expansion of the flux rope and also lift it to an altitude high enough to trigger the torus instability that eventually results in a fast CME \citep{ChengXin2013b, ChengXin2013a}. The eruption of the CME greatly changes the magnetic field structure in the flare region, which further speeds up the magnetic reconnection and enhances the energy release rate.\par 

As the flare evolved into the impulsive phase, the X-ray radiation experienced a sharp increase (Figure 1(a)). The 25--50 keV HXR flux, as well as the derivative of GOES 1.0--8.0 {\AA} flux, had a peak after the CME eruption. Unfortunately, we encountered a RHESSI night from 05:33 to 06:11 UT, and thus lost the RHESSI data at the SXR peak of about 05:58 UT, as estimated from the GOES 1.0--8.0 {\AA} light curve (see Figure 1(a)). After the SXR peak, the flare experienced a long gradual phase that lasted more than 8 hours. The integrated flux of the flare region (Figure 2) in six EUV passbands of AIA (94 {\AA}, 131 {\AA}, 171 {\AA}, 193 {\AA}, 211 {\AA}, and 335 {\AA}) is shown in Figure 1(b). During the initial rise phase and the impulsive phase, the most obvious increase appeared in the 94 {\AA} and 131 {\AA} flux, indicating that a large amount of plasma was heated to a temperature of $\sim$ 10 MK. Obviously, the increases in the 94 {\AA} and 131 {\AA} flux underwent two stages, probably corresponding to the slow reconnection in the initial rise phase and the later fast reconnection associated with the impulsive acceleration of the CME, respectively \citep{ChengXin2013b, ChengXin2013a}. It is interesting that during the rapid increases in the flux of these two high temperature passbands, the flux increases in other low temperature passbands are no more than twice, or even a slight dimming appears after the eruption of the CME. The dimming is synchronous with the most rapidly growing stage of the 94 {\AA} and 131 {\AA} flux, indicating a direct consequence of the heating of cool plasma.         

 \section{DATA ANALYSIS AND THE DEM METHOD}
 
In this work, we choose six AIA EUV passbands, 94 {\AA} (Fe X, Fe XVIII), 131 {\AA} (Fe VIII, Fe XX, Fe XXIII), 171 {\AA} (Fe IX), 193 {\AA} (FeXII, FeXXIV), 211 {\AA} (Fe XIV), and 335 {\AA} (Fe XVI), to calculate the DEM of the flare region as shown in Figure 2. We use the data of AIA from 03:00 to 14:00 UT, with a cadence of 12 s and a spatial resolution of 1.2$\arcsec$. The AIA has two observation modes with different exposure times when observing the flare. Part of the data have a routine exposure time, but losing some information in the loop top and footpoints, especially in the 131 {\AA} and 193 {\AA} images, owing the saturation effect. However, longer exposure time can lead to a better signal-to-noise ratio, especially in some faint regions, such as the reconnection site to which the DEM method can be well applied. The remainder of the data were observed with a reduced exposure time, which can be better used to study the DEM of the flare loop. \par 

 We first use the program ``aia$\_$prep.pro'' to process the AIA data to align the images from six filters, making sure that the accuracy of alignment is better than 0.6$\arcsec$ \citep{Aschwanden2013a}. Since our purpose is to compute the DEM of a selected area that relies on the emissions at different passbands, the alignment of different images is very important. To reduce the impact of misalignment, we degrade the resolution to 2.4$\arcsec$. Moreover, in order to compute the DEM of different regions of the flare, we choose a series of squares, whose centroids are shown in Figure 8(a), Figure 9(a)-(d), and Figure 10(a)-(d), with the size of the squares being 15$\arcsec$, 6.6$\arcsec$, and 6.6$\arcsec$, respectively. For each square, we use the mean $digital$ $number$ (DN) over it for our DEM calculations.\par

Secondly, we use the program ``xrt$\_$dem$\_$iterative2.pro'' \citep{Weber2004a, Golub2004a, Schmelz2010a, Schmelz2011b, Schmelz2011a, Winebarger2011a, ChengXin2012a} to calculate the DEM. The equation of DEM can be written as \citep{Weber2004a, Golub2004a}:
 \begin{equation} 
 {Q_{ic}}  =  \int DEM(T) I_c (T) \mathrm{d}T .
 \end{equation}
The subscript `$c$' donates different channels, and the subscript `$i$' donates the particular pixel in the observations. $Q_{ic}$ is the observational DN (per second), and $I_c(T)$ is the response function of a specific channel. The forward modeling approach first assumes an initial DEM and applies it to Eq. (1) to calculate the predicted observational flux for each channel. It then adjusts the DEM distribution until the predicted fluxes are the closest to the observational ones. This optimal process is done with the mpfit routines provided by Craig B. Markwardt\footnote[1]{http://www.physics.wisc.edu/$\sim$craigm/idl/}. These routines are based on the MINPACK-1 routines\footnote[2]{http://www.netlib.org/minpack/} using a non-linear least-squares method. The uncertainties in the DEM results mainly arise from the statistical noise, the uncertainties of the response functions, and the determining of the foreground and background. Please refer to \cite{ChengXin2012a} for details.\par

The temperature range in our computations is 5.5$\leq$ log${T}\leq$ 7.5, which is mainly determined by the temperature response curves of six AIA filters. In order to characterize the overall thermal property, we define a DEM-weighted temperature and a total emission measure as follow: 
 \begin{equation} 
 T_{\mathrm{mean}} = \frac{ \int_{T_{\mathrm{min}}}^{T{\mathrm{max}}} DEM(T) T \mathrm{d}T} {\int_{T {\mathrm{min}}}^{T{\mathrm{max}}} DEM(T) \mathrm{d}T}    
 \end{equation}\par
 and
 \begin{equation} 
 EM  =  \int _{T_{\mathrm{min}}} ^{T_{\mathrm{max}}} DEM(T) \mathrm{d}T.
 \end{equation}
 
As seen from Figure 4, the emission measure distribution (EMD) is well constrained in the temperature range of 5.7$\leq$LogT$\leq$7.3; so we integrate the DEM over this temperature range to calculate the temperature and EM. In the following sections, the temperature and the EM refer to the DEM-weighted temperature and the total emission measure, respectively.\par

 \section{DEM OF THE FLARE REGION}
 Using the method in \S3, we calculate the DEM of the whole flare region, the reconnection region, the flare loop and the cusp-shaped structure. For each region, we derive the distributions and evolutions of the temperature and EM. \par
 
\subsection{\emph{Overall Property of the Flare}}
Applying the DEM analysis to the whole flare region, we obtain a time series of two-dimensional images of temperature and EM. The snapshots of the flare evolution are shown in Figure 2, in which the four rows correspond to the 131 {\AA}, 193 {\AA}, temperature, and EM images, respectively (the complete evolution movies are available in online materials). \par
  
 When the flare evolved into the initial rise phase, in the 131 {\AA} image (Figure 2(a)), we see that the flare loop structure, especially the northern part of the loop, did not fully appear at that time. The temperature and EM maps show that the flare loop top has already been a structure of high temperature and high density in that stage. In the following time, the temperature and the EM of the loop top kept increasing, and the flux rope began to rise and expand, and finally erupted (the second column). We can see that the upward moving CME kept a high-temperature and high-density structure even when it reached the upper corona (Fig. 2(j) and 2(n)). When the SXR peaked (the third column), we find an extended structure above the loop visible in the 131 {\AA} and 193 {\AA} images but not in other low temperature channels. This implies that the temperature there may be super-high, since the 193 {\AA} filter has a high-temperature response \citep{LiuWei2013a}. This hot region is further confirmed in the temperature map, which is located above the loop top and roughly coincides with the extended structure in the 193 {\AA} image. The temperature of this region can reach as high as $\sim$ 16 MK, which is comparable to that derived from GOES and RHESSI by \cite{LiuWei2013a} and \cite{Battaglia2013a}. However, the highest EM appears underneath the hottest region, with a maximum value of $\sim$ $2.8\times10^{30}$ $\mathrm{cm}^{-5}$. At this moment, both the distributions of EM and 131 {\AA} intensity are apparently asymmetric along the flare loops. In the gradual phase (the fourth column), new flare loops are still continuously formed, though, at a relatively low rate. The most notable feature is that a cusp-shaped structure appears above the flare loop, which can also be clearly seen in the temperature and EM images.\par

To know the overall behavior of the flare, we calculate the average DN over the whole flare region and further derive its DEM. In Figure 3(a), we show the selected area that is, on one hand, big enough to cover the whole flare all the time, and on the other hand, small enough to avoid the contamination from the surrounding structures. Figure 4 shows the Loci curves of the whole flare region at 05:20 UT (about HXR peak), from which we can see that the AIA six passbands have a good constraint on the EMD between Log $T$=5.7 and Log $T$=7.3. Besides, the results from 100 Monte Carlo simulations show that the calculation errors are relatively small. The root mean squares of 100 Monte Carlo simulations are regarded as the uncertainties of the temperature and EM  for the flare region as shown in Figure 3(c) and 3(d).\par

The time evolution of the DEM curve is shown in Figure 3(b). We can see that in an earlier stage, i.e., from 03:00 to 05:10 UT, the DEM of the flare region changed slowly. The abrupt change in DEM curves occurred after 05:10 UT due to the eruption of the flux rope. Around the SXR peak, a double-peak distribution in the DEM curve appeared. The main peak lies at about 10 MK with a DEM value of $\sim$ $4\times10^{22}$ $\mathrm{cm}^{-5}$ $\mathrm{K}^{-1}$. The secondary peak appears at about 1.5 MK, where the DEM is six times lower than that of the main peak. Such a double-peaked temperature distribution is similar to that in \cite{Krucker2013a}, who have also derived temperature of the loop top for this flare using a different DEM inversion code by \cite{Hannah2012a}. It is supposed that the lower temperature peak corresponds to the relatively cool plasma that is superimposed on the hot plasma along the line of sight. There is, however, no doubt that the higher temperature peak is caused by the heating of the plasma when the flare occurred. In Figure 3(c), we show the temperature and EM evolutions of the whole flare region. The three vertical dash lines correspond to the moment of the CME eruption, the HXR peak, and the SXR peak, respectively. The figure shows that the temperature had a rapid increase at the beginning of the flare when the EM increases slowly; a substantial increase in EM appeared at about 04:40 UT. The eruption of the flux rope had a critical effect on the flare thermodynamics. Once the flux rope erupted, the temperature and the EM of the flare sharply increased. The former peaked about ten minutes after the HXR peak, which commenced the start of the fastest growing stage of the EM. The EM reached its maximum more than twenty minutes after the SXR peak. In the gradual phase, the temperature decayed very slowly. About five hours after the SXR peak, the temperature was still as high as $\sim$ 6.5 MK, indicating that continuous heating may last into the later stage.\par
 
To view more clearly the thermodynamics of the flare plasma, we divide the temperature range into three intervals: the low temperature (5.7$\leq$ log ${T}\leq$ 6.0), the medium temperature (6.1$\leq$ log ${T}\leq$ 6.7), and the high temperature (6.8$\leq$ log ${T}\leq$ 7.3) ones, and then integrate the DEM in each interval. The results show that, after the beginning of the flare ($\sim$ 04:17 UT), the high temperature plasma had a rapid increase, but the plasma with low and medium temperatures did not change so much. As shown in Figure 2(j), the hot plasma is mainly located below and above the reconnection site in the initial rise phase. Thus, we infer that the initial increase of the high temperature plasma may be directly from the reconnection outflows. As the reconnection continues, the pre-heated plasma is further heated in the outflow regions by mechanisms such as turbulence and plasma waves \citep{LiuWei2013a}. After the CME eruption, the lower temperature plasma suddenly decreased at a fast rate until the HXR peak, then it began to increase again. In about forty minutes after the SXR peak, the EM in different temperatures evolved without any significant change. It is known that without further heating, the plasma should cool down. Thus this ``gentle" phase may imply that some kind of heating may keep working compensating for the cooling effect.\par 
   
 \subsection{\emph{Magnetic Reconnection Region}}

The magnetic reconnection region is the central part of a flare since it is where the magnetic energy is converted to other forms of energy. However, it is still debated where the energy is most effectively released to thermal energy of plasma and kinetic energy of accelerated particles. \cite{LiuWei2013a} thought it happens in the outflow region rather than the reconnection site itself. To quantitatively check this argument, we study in detail the DEM distribution of the reconnection region.

 \subsubsection{Plasma heating in pre-flare phase}
 
In Figure 5(a)-(h), we show the temperature and EM maps at four selected times, which can reveal the initial evolution of the flare. We also select three regions located at the southern footpoint, the loop top, and the northern footpoint, respectively and plot the evolution curves of the temperature and the EM in these regions in Figure 5(i)-(j).\par

 In the pre-flare phase (before 04:17 UT), the temperature ($\sim$ 4.5 MK) and the EM ($\sim2.5\times10^{28}$ $\mathrm{cm}^{-5}$) of the loop top were already higher than the surroundings. Actually, we find that both of them began to increase at about 04:10 UT (see Figure 5(i)-(j)). Meanwhile, we did not detect substantial change of flux in low temperature passbands as shown in Figure 1(b). Due to the resolution limit of AIA, we can not determine the heating process of the hot plasma at this stage. At about 04:16 UT, there are two high-temperature and high-density regions, located at the loop top and the southern footpoint, respectively. In the following 12 minutes, the southern part of the loop was filled gradually with hot plasma. During this period, however, the northern part of the loop was not completely discernable. Above the reconnection region, there also existed a high-temperature and high-density structure, which could be caused by the upward reconnection outflow.\par

 \subsubsection{Reconnection region in initial rise phase}
 
Previous researches showed that the CME eruption has a close relationship with the flare energy release \citep{zhangJie2001a, Temmer2008a, ChengXin2010a}. Before the CME eruption, the active region is relatively stable, and the SXR emission increases at a smaller rate compared with that after the CME eruption. We plot the distribution of the temperature and EM of the flare region in Figure 6. The possible site, where the magnetic reconnection takes place, is between the flare loop top and the flux rope, as inferred from the configuration of this area \citep{LiuWei2013a}. We conjecture that it is located close to the middle RHESSI HXR source, as noted below (see Figure 6(b)). In about half an hour from the start of the flare ($\sim$ 04:17 to 04:47 UT), the flux rope experienced an expansion and a rising motion. Meanwhile, the temperature and EM of these two structures kept increasing gradually. We overplot the RHESSI X-ray sources on the 131{\AA}, temperature, and EM images. The RHESSI images are reconstructed using the clean algorithm, and with the time interval for integration being from 04:45 to 04:50 UT. The detectors that we choose are 3F, 5F, 6F, 7F, 8F, and 9F, and the energy bands are 6--10 keV and 10--15 keV. We do 600 iterations for each energy band.\par

As shown in Figure 6(b), we find that the temperature of the reconnection site is not the highest. The two locally hottest regions, with temperatures of 9-10 MK, are located in the loop top and the flux rope. In Figure 6(c), the EM map does not show a high-density structure near the reconnection site either. The two high-density regions above and below the reconnection site have a good correspondence to the two hot regions in the temperature map. Besides, there are two X-ray sources correlating well with these two high-density and high-temperature regions. As \cite{LiuWei2013a} mentioned, a variety of possible acceleration and heating mechanisms may work in the outflow region like turbulence, fast-mode shocks, first-order Fermi, and betatron mechanisms. They thought that the most effective particle acceleration and plasma heating take place outside of the current sheet. Our results here are basically consistent with that of \cite{LiuWei2013a}. \par
 
 As long as the magnetic reconnection is underway, the loop top and the flux rope keep being heated. The temperature rises and the magnetic configuration changes gradually. Therefore, a local high pressure region can be formed near the bottom of the flux rope. Such a pressure can provide an upward force on the flux rope, which keeps being enhanced with continuous heating. On the other hand, the magnetic reconnection reconfigures the magnetic field, and its constrained force on the flux rope decreases. These two factors increase the net upward force to accelerate the CME eruption impulsively.\par
 
It is interesting that besides the loop top source and the coronal source, there seems to exist a third source in our reconstructed RHESSI HXR images. As shown in Figure 6(b), this source is located between the loop top and the coronal sources, correlating well with the possible reconnection site. To our knowledge, such a HXR source has not been reported before. If this is true, it implies some local heating at the reconnection site, possibly corresponding to some fine structures like the magnetic islands within the current sheet. However, we cannot rule out the possibility that the third source is just an artifact due to the uncertainty of the HXR image reconstruction. We should check this point using more events in the future.\par

  \subsubsection{Cusp-shaped structure in the gradual phase}
  
In the gradual phase, a cusp-shaped structure is formed as shown in Figure 7. Above this, there exists a long sheet structure that was identified as a vertical current sheet (VCS) by \cite{LiuRui2013a}. We calculate the temperature and the EM of this region and show the results in Figure 7. Because the cusp-shaped structure is located in the upper corona where the EUV radiation is faint, we degrade the resolution of the AIA data to 4.8$\arcsec$ for a better signal-to-noise ratio.\par

The cusp-shaped structure is an important component of the CSHKP model. It is taken as a signature of magnetic reconnection. From Figure 7(c)-(d), one can find that the cusp-shaped structure is of high temperature and high density. We put two slices in this region: one is along the sheet structure, and the other is perpendicular to the sheet structure. The temperature and EM distributions along these two slices are shown in Figure 7(e) and 7(f), and the results are smoothed with a width of 2 points. Along slice 1, both the temperature and the EM decrease with height. The EM decreases to about 16\% of the value at the lowermost point. By comparison, the temperature along the sheet just has a slight decrease, say, from 8.7 to 7.2 MK. If we assume that the depth of the sheet along the line of sight is uniform, then the pressure gradient of the plasma along the sheet is upward. However, it is found that the plasma moves downward along the sheet \citep{LiuWei2013a, LiuRui2013a}; so there must be other downward forces exerting on the plasma like gravity and magnetic force. On the northern side of the sheet structure, the most notable feature is a sharp change in both the temperature and the EM distribution curves (Figure 8(f)), which may be evidence of a slow mode shock produced by magnetic reconnection. That still works in the gradual phase. \par

 \subsection{\emph{Dynamics of the Flare Loop}}
 
 \subsubsection{Temperature and EM in the loop top}

The flare loop top is of special interest since it is usually the closest loop part to the reconnection site. Impinging of the reconnection outflow with the flare loop may cause a significant heating. To study this process in detail, we identify two representative points, the loop top point and the hottest point. We adopt the weighted center of 131 {\AA} emission ($\ge$ 80\% the maximum DN) as the loop top point and the weighted center of temperature ($\ge$ 95\% the maximum temperature) as the hottest point, respectively. The trajectories of the loop top point and the hottest point are plotted in Figure 8(a) by colored crosses and triangles, respectively. Centered on these points, we set a series of boxes with the size of $\sim$ $15\arcsec$. The mean AIA DN value in each box is used in our calculations.\par

It is clear that the hottest region always lies higher than the loop top, as shown in Figure 8(b). Both the loop top and the hottest region underwent an up-down-up motion. This motion is consistent with the motion of RHESSI loop top X-ray source described by \cite{LiuWei2013a}, and it may be partly due to the up-down-up motion of the reconnection site \citep{LiuWei2013a}. In the initial rise phase, the EM in the loop top rises monotonically at a relatively slow rate (Figure 8(c)). After the CME eruption, it begins to rise at a faster rate, and reaches a maximum value of $\sim$ $2.5\times10^{30}$ $\mathrm{cm}^{-5}$ at about 06:10 UT ($\sim$12 minutes after the SXR peak). After that, the EM decreases monotonously in the gradual phase. As for the temperature in the loop top, it has a maximum value when the HXR peaked. In the gradual phase, it undergoes a decay as shown in Figure 8(c).\par

For comparison, we also calculate the EM in the flare hottest region (shown in Figure 8(d)), and find that it stopped increasing when the HXR peaked, reaching a maximum value of $\sim$ $3.5\times10^{29}$ $\mathrm{cm}^{-5}$, which is far less than that in the loop top. It is known that, in the impulsive phase, the loop top part is densified mainly through the chromospheric evaporation, which is restricted in the flare loop. Since the hottest region is obviously outside of (above) the flare loop, the chromospheric evaporation has less effects on the EM in the hottest region than that in the loop top. Evolution of the EM of the hottest region may mostly be related to the magnetic reconnection process. During the whole impulsive phase, we find a more substantial increase of the EM in the loop top than that in the hottest region, highlighting the importance of chromospheric evaporation in increasing the plasma in the flare loop.\par

 \subsubsection{Temperature and EM along the flare loop}
 
 The distributions of temperature and EM along the loop are critical to diagnose the dynamics of the flare. As the flare loop observed by different filters may have different shapes, here we choose the flare loops shaped in the 131 {\AA} images for our study. In principal, the loops in the 131 {\AA} images are hot and believed to be formed by the newly reconnected fields. We discretize the 131 {\AA} loops into a series of grids, as shown in Figure 9. For each grid point, we use the mean DN of a $6.6\arcsec$$\times$$6.6\arcsec$ square centered on the point. The deduced temperature and EM distributions along the flare loop are shown in Figure 9(e)-(h). The four columns correspond to four stages in the flare evolution (the initial rise phase, the HXR peak, the SXR peak, and the gradual phase). The error bars are determined from the results of 100 Monte Carlo simulations. It may be noted that what we analyze here are the newly formed loops at different times rather than the same loop. Moreover, the loops as depicted  by the 131 {\AA} emission may deviate somewhat from the real flare loops constrained by magnetic fields. However, it has no direct influence on the EM analysis of the loops but at most a small influence on the loop dynamics ($\S$ 4.3.3). Since such a geometrical difference should be small, we think that it does not alter our results qualitatively.\par
    
At all the four selected times, the hottest part of the loop is always the loop top with temperatures varying from 8 to 14 MK. From the loop top to the footpoints, the temperature decreases monotonically; in the footpoints, the temperature varies in the range of 6 to 10 MK, which is consistent with the result of \cite{Graham2013a}. The change of temperature at different times is not very significant, with the biggest increase being less than twice. By contrast, the change of the EM is much larger. For example, the EM is about $10^{29}$ $\mathrm{cm}^{-5}$ in the northern footpoint in the initial rise phase, and increases to more than $10^{30}$ $\mathrm{cm}^{-5}$ at the SXR peak. At about 05:00 UT (the initial rise phase), the distribution of EM resembles the distribution of temperature: the EM decreases monotonically from the loop top to the footpoints. While after the eruption of the CME, as shown in Figure 9(f) and 9(g), the most significant increase of the EM occurs in the northern footpoint, where the EM rises from $9.0\times10^{28}$ to $1.0\times10^{30}$ $\mathrm{cm}^{-5}$. As a result, a high EM region is formed in the northern footpoint. At that time, the highest EM region is still at the loop top; from the loop top to the loop footpoints, the EM first decreases and then increases. At the SXR peak, the EM along the flare loop is extremely asymmetric that the value in the northern part of the loop is significantly higher than that in the southern part. However, in the gradual phase, this asymmetric distribution in EM becomes less obvious. About 140 minutes after the SXR peak (Figure 9(h)), we find that the temperature in the loop top is still pretty high, i.e., 9 MK. It implies that there might exist a continuous heating process in the loop top, compensating to some extent the cooling effect.\par

The high EM region in the northern footpoint is probably caused by the plasma injection. This plasma injection is believed to be a result of chromospheric evaporation. The asymmetric EM at the two footpoints could be due to asymmetric magnetic fields and the mirroring effect, which result in different loss-cones for precipitating electrons and collisional heating to the chromospheric plasma. This picture appears to be consistent with the asymmetric HXR fluxes at the two footpoints with the northern one being stronger, as shown in \cite{LiuRui2013a}, although additional contribution to the HXR asymmetry could come from the partial limb occultation of the southern footpoint \citep{Battaglia2013a}. \par

\subsubsection{Dynamics of the flare loop}

With the deduced temperature and EM, we can analyze the dynamics of the flare loop, assuming that the 131 {\AA} loops can roughly trace magnetic field lines, as discussed above. We consider the one-dimensional problem, i.e., plasma motion constrained along the flare loop. The plasma is taken as frictionless and incompressible. In the direction parallel to the flare loop, the forces acting on the plasma are pressure ($F_{\mathrm{p}}$) and gravity ($F_{\mathrm{g}}\cos\theta$, $\theta$ is the angle between the loop segment and gravity). Furthermore, we assume that the plasma consists of hydrogen and helium, with an abundance ratio $n_{\mathrm{H}} $ : $n_{\mathrm{He}}$ = 10 : 1. In the corona, the atoms are fully ionized; therefore, $n_{\mathrm{e}} = n_{\mathrm{H}} + 2 n_{\mathrm{He}}$. \par

 The motion of plasma is controlled by the one-dimensional hydrodynamic equation:
 \begin{equation} 
\rho \frac{\mathrm{d} u}{\mathrm{d} t} = \rho F - \nabla P,
 \end{equation}
 where $F = F_{\mathrm{g}} \cos\theta$, $\rho= n_{\mathrm{H}} \mathrm{m_H} + n_{\mathrm{He}} \mathrm{m_{He}}$, and $P=(n_{\mathrm{e}} + n_{\mathrm{H}} + n_{\mathrm{He}}) \mathrm{k} T$.
 As in $\S$ 4.3.2, we choose a set of points along the 131 {\AA} flare loop for calculations. We adopt a difference method with two-order accuracy; thus, every three adjacent points constitute a calculation unit. We first calculate the DEM along the flare loop, using a mean DN value over a $6.6\arcsec\times6.6\arcsec$ square centered on each point. Then we calculate the net force exerted on the plasma per gram (the acceleration) for four selected times, corresponding to the initial rise phase, the CME eruption, the beginning of the impulsive phase, and the impulsive phase, respectively, as shown in Figure 10(a)-(d).\par
 
The results are shown in Figure 10(e). The figure shows that the acceleration is downward in most parts of the flare loop. From the loop top to the footpoints, the absolute value of acceleration first increases and reaches a maximum value at some mid-points in the loop legs, implying the gradually increasing gravity component along the loop; however, it then decreases towards the footpoints, implying that the gravity is largely compensated by the upward driving force, which could be related to chromospheric evaporation. In addition, the distribution of the acceleration along the loop changes with time, especially in the northern footpoint. As seen from Figure 10(e), the acceleration in the northern footpoint is downward initially, and then it decreases. At about 05:08 UT (the moment of the CME eruption), a quasi-hydrostatic state is reached in the northern footpoint. In the next 16 minutes, an upward acceleration appears in the northern footpoint, which reaches a maximum value of about $5\times10^{5}$ $\mathrm{cm}$ $\mathrm{s^{-2}}$ ($\sim$ 20 times the gravitational acceleration at the solar surface). This is a clear signature of chromospheric evaporation. In contrast, we do not see any obvious change of the acceleration in the southern footpoint. A possible explanation is that the electron heating in the southern footpoint is relatively weak, as noted above. \par  

The chromosphere evaporation plays an important role in filling in the flare loop with hot plasma. The increase of the EM in the flare region can be mostly attributed to this process. Chromospheric evaporation has been extensively studied mainly by observing the plasma flow velocity characterized by Doppler shifts in EUV and SXR lines \citep{Antonucci1983a, Ding1996a, Brosius2004a, Doschek2005a, Milligan2009a, LiYing2011a}. In this work, however, we provide evidence of chromospheric evaporation in the aspect of hydrodynamics based on DEM analysis.

 \section{Summary and Conclusions}
In this work, we analyze an M7.7 limb flare on 19 July 2012 using the high resolution EUV data observed by \emph{SDO}/AIA. By applying a DEM method, we obtain the quantitative distributions of temperature and EM of the flare region including the flare loop and the reconnection site. The main results are summarized as follows.\par
\begin{enumerate}
\item{At the beginning of the flare, a significant amount of hot plasma ($\sim$ 5 MK) appeared in the top of the flare loop. As the reconnection continues, some heating mechanisms (such as turbulence or plasma waves) further heat the plasma to a higher temperature of $\ge$ 10 MK in the outflow regions, consistent with the results of \cite{LiuWei2013a}.}

\item{Along the flare loop, the temperature and the EM are the highest in the loop top. From the loop top to the footpoints, the temperature and the EM decrease monotonically in the initial phase. However, the EM in the northern footpoint has a rapid increase during the impulsive phase, which is regarded as evidence of chromospheric evaporation. As a result, the net force exerted on the plasma in the northern footpoint changes its direction from downward to upward. Meanwhile, the chromospheric evaporation in the southern footpoint is weak, probably due to the asymmetry of the magnetic topology.} 
 
 \item{The cusp-shaped structure above the flare loop in the gradual phase is a high-temperature and high-density structure. Above that, there is an elongated structure, probably corresponding to the current sheet. Across the current sheet, there exists a sharp change of temperature and EM, in particular at the northern side, which is probably a signature of a slow MHD shock.}
\end{enumerate}\par
The above studies and results on the spatially resolved DEM provide an example and clues in learning where and when the energy is released during a flare. In the future, we need to study more events toward a better understanding of the flare energetics and thermal dynamics.

\acknowledgements The authors thank W. Liu and the referee for many valuable comments that led to an improvement of the paper. SDO is a mission of NASA's Living With a Star Program. J.Q.S., X.C., and M.D.D. are supported by NSFC under grants 10933003, 11303016, 11373023, 11203014, and NKBRSF
under grants 2011CB811402 and 2014CB744203.

\begin{thebibliography}{67}
\expandafter\ifx\csname natexlab\endcsname\relax\def\natexlab#1{#1}\fi

\bibitem[{{Antonucci} \& {Dennis}(1983)}]{Antonucci1983a}
{Antonucci}, E., \& {Dennis}, B.~R. 1983, \solphys, 86, 67

\bibitem[{{Aschwanden} {et~al.}(2013){Aschwanden}, {Boerner}, {Schrijver}, \&
  {Malanushenko}}]{Aschwanden2013a}
{Aschwanden}, M.~J., {Boerner}, P., {Schrijver}, C.~J., \& {Malanushenko}, A.
  2013, \solphys, 283, 5

\bibitem[{{Battaglia} \& {Kontar}(2012)}]{Battaglia2012a}
{Battaglia}, M., \& {Kontar}, E.~P. 2012, \apj, 760, 142

\bibitem[{{Battaglia} \& {Kontar}(2013)}]{Battaglia2013a}
---. 2013, ArXiv e-prints

\bibitem[{{Bornmann}(1985)}]{Bornmann1985a}
{Bornmann}, P.~L. 1985, \solphys, 102, 111

\bibitem[{{Brosius} \& {Phillips}(2004)}]{Brosius2004a}
{Brosius}, J.~W., \& {Phillips}, K.~J.~H. 2004, \apj, 613, 580

\bibitem[{{Carmichael}(1964)}]{Carmichael1964a}
{Carmichael}, H. 1964, NASA Special Publication, 50, 451

\bibitem[{{Cheng}(1977)}]{Cheng1977a}
{Cheng}, C.-C. 1977, \solphys, 55, 413

\bibitem[{{Cheng} {et~al.}(2010){Cheng}, {Ding}, {Guo}, {Zhang}, {Jing}, \&
  {Wiegelmann}}]{ChengXin2010a}
{Cheng}, X., {Ding}, M.~D., {Guo}, Y., {Zhang}, J., {Jing}, J., \&
  {Wiegelmann}, T. 2010, \apjl, 716, L68

\bibitem[{{Cheng} {et~al.}(2013{\natexlab{a}}){Cheng}, {Zhang}, {Ding},
  {Olmedo}, {Sun}, {Guo}, \& {Liu}}]{ChengXin2013b}
{Cheng}, X., {Zhang}, J., {Ding}, M.~D., {Olmedo}, O., {Sun}, X.~D., {Guo}, Y.,
  \& {Liu}, Y. 2013{\natexlab{a}}, \apjl, 769, L25

\bibitem[{{Cheng} {et~al.}(2012){Cheng}, {Zhang}, {Saar}, \&
  {Ding}}]{ChengXin2012a}
{Cheng}, X., {Zhang}, J., {Saar}, S.~H., \& {Ding}, M.~D. 2012, \apj, 761, 62

\bibitem[{{Cheng} {et~al.}(2013{\natexlab{b}}){Cheng}, {Ding}, {Guo}, {Zhang},
  {Vourlidas}, {Liu}, {Olmedo}, {Sun}, \& {Li}}]{ChengXin2013a}
{Cheng}, X., {et~al.} 2013{\natexlab{b}}, ArXiv e-prints

\bibitem[{{Denton} \& {Feldman}(1984)}]{Denton1984a}
{Denton}, R.~E., \& {Feldman}, U. 1984, \apj, 286, 359

\bibitem[{{Dere} {et~al.}(1974){Dere}, {Horan}, \& {Kreplin}}]{Dere1974a}
{Dere}, K.~P., {Horan}, D.~M., \& {Kreplin}, R.~W. 1974, \solphys, 36, 459

\bibitem[{{Dere} {et~al.}(1977){Dere}, {Horan}, \& {Kreplin}}]{Dere1977a}
---. 1977, \apj, 217, 976

\bibitem[{{Dere} {et~al.}(1979){Dere}, {Widing}, {Mason}, \&
  {Bhatia}}]{Dere1979a}
{Dere}, K.~P., {Widing}, K.~G., {Mason}, H.~E., \& {Bhatia}, A.~K. 1979, \apjs,
  40, 341

\bibitem[{{Ding} \& {Fang}(1996)}]{Ding1996a}
{Ding}, M.~D., \& {Fang}, C. 1996, \aap, 314, 643

\bibitem[{{Doschek} {et~al.}(1981){Doschek}, {Feldman}, {Landecker}, \&
  {McKenzie}}]{Doschek1981a}
{Doschek}, G.~A., {Feldman}, U., {Landecker}, P.~B., \& {McKenzie}, D.~L. 1981,
  \apj, 249, 372

\bibitem[{{Doschek} \& {Warren}(2005)}]{Doschek2005a}
{Doschek}, G.~A., \& {Warren}, H.~P. 2005, \apj, 629, 1150

\bibitem[{{Duijveman}(1983)}]{Duijveman1983a}
{Duijveman}, A. 1983, \solphys, 84, 189

\bibitem[{{Feldman} {et~al.}(1996{\natexlab{a}}){Feldman}, {Doschek}, \&
  {Behring}}]{Feldman1996b}
{Feldman}, U., {Doschek}, G.~A., \& {Behring}, W.~E. 1996{\natexlab{a}}, \apj,
  461, 465

\bibitem[{{Feldman} {et~al.}(1996{\natexlab{b}}){Feldman}, {Doschek},
  {Behring}, \& {Phillips}}]{Feldman1996a}
{Feldman}, U., {Doschek}, G.~A., {Behring}, W.~E., \& {Phillips}, K.~J.~H.
  1996{\natexlab{b}}, \apj, 460, 1034

\bibitem[{{Fletcher}(1995)}]{Fletcher1995a}
{Fletcher}, L. 1995, \aap, 303, L9

\bibitem[{{Fletcher}(1996)}]{Fletcher1996a}
---. 1996, \aap, 310, 661

\bibitem[{{Fletcher} \& {Martens}(1998)}]{Fletcher1998a}
{Fletcher}, L., \& {Martens}, P.~C.~H. 1998, \apj, 505, 418

\bibitem[{{Golub} {et~al.}(2004){Golub}, {Deluca}, {Sette}, \&
  {Weber}}]{Golub2004a}
{Golub}, L., {Deluca}, E.~E., {Sette}, A., \& {Weber}, M. 2004, in Astronomical
  Society of the Pacific Conference Series, Vol. 325, The Solar-B Mission and
  the Forefront of Solar Physics, ed. T.~{Sakurai} \& T.~{Sekii}, 217

\bibitem[{{Graham} {et~al.}(2013){Graham}, {Hannah}, {Fletcher}, \&
  {Milligan}}]{Graham2013a}
{Graham}, D.~R., {Hannah}, I.~G., {Fletcher}, L., \& {Milligan}, R.~O. 2013,
  \apj, 767, 83

\bibitem[{{Hahn} {et~al.}(2011){Hahn}, {Landi}, \& {Savin}}]{Hahn2011a}
{Hahn}, M., {Landi}, E., \& {Savin}, D.~W. 2011, \apj, 736, 101

\bibitem[{{Hannah} \& {Kontar}(2012)}]{Hannah2012a}
{Hannah}, I.~G., \& {Kontar}, E.~P. 2012, \aap, 539, A146

\bibitem[{{Hirayama}(1974)}]{Hirayama1974a}
{Hirayama}, T. 1974, \solphys, 34, 323

\bibitem[{{Holman}(1985)}]{Holman1985a}
{Holman}, G.~D. 1985, \apj, 293, 584

\bibitem[{{Horan}(1971)}]{Horan1971a}
{Horan}, D.~M. 1971, \solphys, 21, 188

\bibitem[{{Hudson}(1972)}]{Hudson1972a}
{Hudson}, H.~S. 1972, \solphys, 24, 414

\bibitem[{{Kahler} {et~al.}(1970){Kahler}, {Meekins}, {Kreplin}, \&
  {Bowyer}}]{Kahler1970a}
{Kahler}, S.~W., {Meekins}, J.~F., {Kreplin}, R.~W., \& {Bowyer}, C.~S. 1970,
  \apj, 162, 293

\bibitem[{{Kopp} \& {Pneuman}(1976)}]{Kopp1976a}
{Kopp}, R.~A., \& {Pneuman}, G.~W. 1976, \solphys, 50, 85

\bibitem[{{Krucker} \& {Battaglia}(2013)}]{Krucker2013a}
{Krucker}, S., \& {Battaglia}, M. 2013, in AAS/Solar Physics Division Meeting,
  Vol.~44, AAS/Solar Physics Division Meeting, 402.04

\bibitem[{{Landini} \& {Monsignori Fossi}(1979)}]{Landini1979a}
{Landini}, M., \& {Monsignori Fossi}, B.~C. 1979, \aap, 72, 171

\bibitem[{{Li} \& {Ding}(2011)}]{LiYing2011a}
{Li}, Y., \& {Ding}, M.~D. 2011, \apj, 727, 98

\bibitem[{{Li} \& {Gan}(2007)}]{Li2007a}
{Li}, Y.~P., \& {Gan}, W.~Q. 2007, Advances in Space Research, 39, 1389

\bibitem[{{Liu}(2013)}]{LiuRui2013a}
{Liu}, R. 2013, \mnras, 434, 1309

\bibitem[{{Liu} {et~al.}(2013){Liu}, {Chen}, \& {Petrosian}}]{LiuWei2013a}
{Liu}, W., {Chen}, Q., \& {Petrosian}, V. 2013, \apj, 767, 168

\bibitem[{{McTiernan} {et~al.}(1993){McTiernan}, {Kane}, {Loran}, {Lemen},
  {Acton}, {Hara}, {Tsuneta}, \& {Kosugi}}]{McTiernan1993a}
{McTiernan}, J.~M., {Kane}, S.~R., {Loran}, J.~M., {Lemen}, J.~R., {Acton},
  L.~W., {Hara}, H., {Tsuneta}, S., \& {Kosugi}, T. 1993, \apjl, 416, L91

\bibitem[{{Melrose}(1995)}]{Melrose1995a}
{Melrose}, D.~B. 1995, \apj, 451, 391

\bibitem[{{Melrose}(1997)}]{Melrose1997a}
---. 1997, \apj, 486, 521

\bibitem[{{Metcalf} {et~al.}(1990{\natexlab{a}}){Metcalf}, {Canfield},
  {Avrett}, \& {Metcalf}}]{Metcalf1990a}
{Metcalf}, T.~R., {Canfield}, R.~C., {Avrett}, E.~H., \& {Metcalf}, F.~T.
  1990{\natexlab{a}}, \apj, 350, 463

\bibitem[{{Metcalf} {et~al.}(1990{\natexlab{b}}){Metcalf}, {Canfield}, \&
  {Saba}}]{Metcalf1990b}
{Metcalf}, T.~R., {Canfield}, R.~C., \& {Saba}, J.~L.~R. 1990{\natexlab{b}},
  \apj, 365, 391

\bibitem[{{Milkey} {et~al.}(1971){Milkey}, {Blocker}, {Chambers}, {Fehlau},
  {Fuller}, \& {Kunz}}]{Milkey1971a}
{Milkey}, R.~W., {Blocker}, N.~K., {Chambers}, W.~H., {Fehlau}, P.~E.,
  {Fuller}, J.~C., \& {Kunz}, W.~E. 1971, \solphys, 20, 400

\bibitem[{{Milligan} \& {Dennis}(2009)}]{Milligan2009a}
{Milligan}, R.~O., \& {Dennis}, B.~R. 2009, \apj, 699, 968

\bibitem[{{O'Dwyer} {et~al.}(2010){O'Dwyer}, {Del Zanna}, {Mason}, {Weber}, \&
  {Tripathi}}]{ODwyer2010a}
{O'Dwyer}, B., {Del Zanna}, G., {Mason}, H.~E., {Weber}, M.~A., \& {Tripathi},
  D. 2010, \aap, 521, A21

\bibitem[{{Patsourakos} {et~al.}(2013){Patsourakos}, {Vourlidas}, \&
  {Stenborg}}]{Patsourakos2013a}
{Patsourakos}, S., {Vourlidas}, A., \& {Stenborg}, G. 2013, \apj, 764, 125

\bibitem[{{Petschek}(1964)}]{Petschek1964a}
{Petschek}, H.~E. 1964, NASA Special Publication, 50, 425

\bibitem[{{Plowman} {et~al.}(2013){Plowman}, {Kankelborg}, \&
  {Martens}}]{Plowman2013a}
{Plowman}, J., {Kankelborg}, C., \& {Martens}, P. 2013, \apj, 771, 2

\bibitem[{{Reeves} \& {Weber}(2009)}]{Reeves2009a}
{Reeves}, K.~K., \& {Weber}, M.~A. 2009, in Astronomical Society of the Pacific
  Conference Series, Vol. 415, The Second Hinode Science Meeting: Beyond
  Discovery-Toward Understanding, ed. B.~{Lites}, M.~{Cheung}, T.~{Magara},
  J.~{Mariska}, \& K.~{Reeves}, 443

\bibitem[{{Schmelz} {et~al.}(2011{\natexlab{a}}){Schmelz}, {Rightmire}, {Saar},
  {Kimble}, {Worley}, \& {Pathak}}]{Schmelz2011b}
{Schmelz}, J.~T., {Rightmire}, L.~A., {Saar}, S.~H., {Kimble}, J.~A., {Worley},
  B.~T., \& {Pathak}, S. 2011{\natexlab{a}}, \apj, 738, 146

\bibitem[{{Schmelz} {et~al.}(2010){Schmelz}, {Saar}, {Nasraoui}, {Kashyap},
  {Weber}, {DeLuca}, \& {Golub}}]{Schmelz2010a}
{Schmelz}, J.~T., {Saar}, S.~H., {Nasraoui}, K., {Kashyap}, V.~L., {Weber},
  M.~A., {DeLuca}, E.~E., \& {Golub}, L. 2010, \apj, 723, 1180

\bibitem[{{Schmelz} {et~al.}(2011{\natexlab{b}}){Schmelz}, {Worley},
  {Anderson}, {Pathak}, {Kimble}, {Jenkins}, \& {Saar}}]{Schmelz2011a}
{Schmelz}, J.~T., {Worley}, B.~T., {Anderson}, D.~J., {Pathak}, S., {Kimble},
  J.~A., {Jenkins}, B.~S., \& {Saar}, S.~H. 2011{\natexlab{b}}, \apj, 739, 33

\bibitem[{{Spicer}(1981{\natexlab{a}})}]{Spicer1981b}
{Spicer}, D.~S. 1981{\natexlab{a}}, \solphys, 71, 115

\bibitem[{{Spicer}(1981{\natexlab{b}})}]{Spicer1981a}
---. 1981{\natexlab{b}}, \solphys, 70, 149

\bibitem[{{Sturrock}(1966)}]{Sturrock1966a}
{Sturrock}, P.~A. 1966, \nat, 211, 695

\bibitem[{{Temmer} {et~al.}(2008){Temmer}, {Veronig}, {Vr{\v s}nak},
  {Ryb{\'a}k}, {G{\"o}m{\"o}ry}, {Stoiser}, \& {Mari{\v
  c}i{\'c}}}]{Temmer2008a}
{Temmer}, M., {Veronig}, A.~M., {Vr{\v s}nak}, B., {Ryb{\'a}k}, J.,
  {G{\"o}m{\"o}ry}, P., {Stoiser}, S., \& {Mari{\v c}i{\'c}}, D. 2008, \apjl,
  673, L95

\bibitem[{{Tsuneta}(1997)}]{Tsuneta1997a}
{Tsuneta}, S. 1997, \apj, 483, 507

\bibitem[{{Voitenko}(1995)}]{Voitenko1995a}
{Voitenko}, Y.~M. 1995, \solphys, 161, 197

\bibitem[{{Voitenko}(1996)}]{Voitenko1996a}
---. 1996, \solphys, 168, 219

\bibitem[{{Voitenko} \& {Goossens}(1999)}]{Voitenko1999a}
{Voitenko}, Y.~M., \& {Goossens}, M. 1999, in ESA Special Publication, Vol.
  448, Magnetic Fields and Solar Processes, ed. A.~{Wilson} \& {et al.}, 735

\bibitem[{{Weber} {et~al.}(2004){Weber}, {Deluca}, {Golub}, \&
  {Sette}}]{Weber2004a}
{Weber}, M.~A., {Deluca}, E.~E., {Golub}, L., \& {Sette}, A.~L. 2004, in IAU
  Symposium, Vol. 223, Multi-Wavelength Investigations of Solar Activity, ed.
  A.~V. {Stepanov}, E.~E. {Benevolenskaya}, \& A.~G. {Kosovichev}, 321--328

\bibitem[{{Winebarger} {et~al.}(2011){Winebarger}, {Schmelz}, {Warren}, {Saar},
  \& {Kashyap}}]{Winebarger2011a}
{Winebarger}, A.~R., {Schmelz}, J.~T., {Warren}, H.~P., {Saar}, S.~H., \&
  {Kashyap}, V.~L. 2011, \apj, 740, 2

\bibitem[{{Zhang} {et~al.}(2001){Zhang}, {Dere}, {Howard}, {Kundu}, \&
  {White}}]{zhangJie2001a}
{Zhang}, J., {Dere}, K.~P., {Howard}, R.~A., {Kundu}, M.~R., \& {White}, S.~M.
  2001, \apj, 559, 452

\end{thebibliography}


\begin{figure*} 
      \vspace{-0.0\textwidth}    
      \centerline{\hspace*{0.00\textwidth}
      \includegraphics[width=1.0\textwidth,clip=]{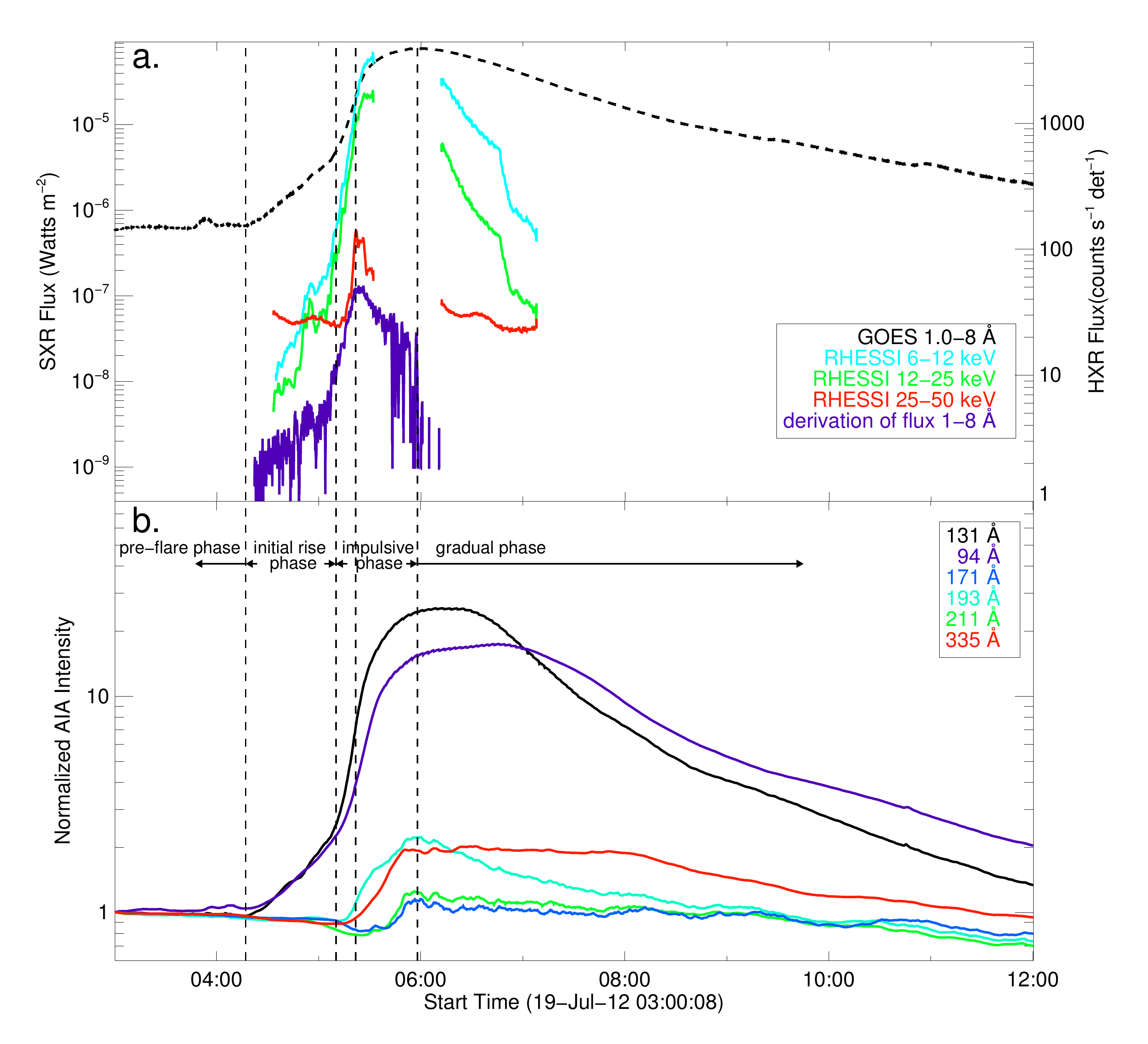}
      }
\caption{ (a) Light curves of GOES 1.0-8.0 {\AA}, RHESSI 6-12 keV, 12-25 keV, and 25-50 keV  for the 2012 July 19 flare. The time derivative of GOES 1.0-8.0 {\AA} flux is also shown. (b) Light curves of AIA 94 {\AA}, 131 {\AA}, 171 {\AA}, 193 {\AA}, 211 {\AA}, and 335 {\AA}  normalized to the initial values at 03:00 UT. The four vertical lines correspond to the start of the flare, the CME eruption, the RHESSI 25-50 keV flux peak, and the SXR peak, respectively.} \label{f1}
\end{figure*}

\begin{figure*} 
     \vspace{-0.0\textwidth}    
     \centerline{\hspace*{0.00\textwidth}
               \includegraphics[width=1.0\textwidth,clip=]{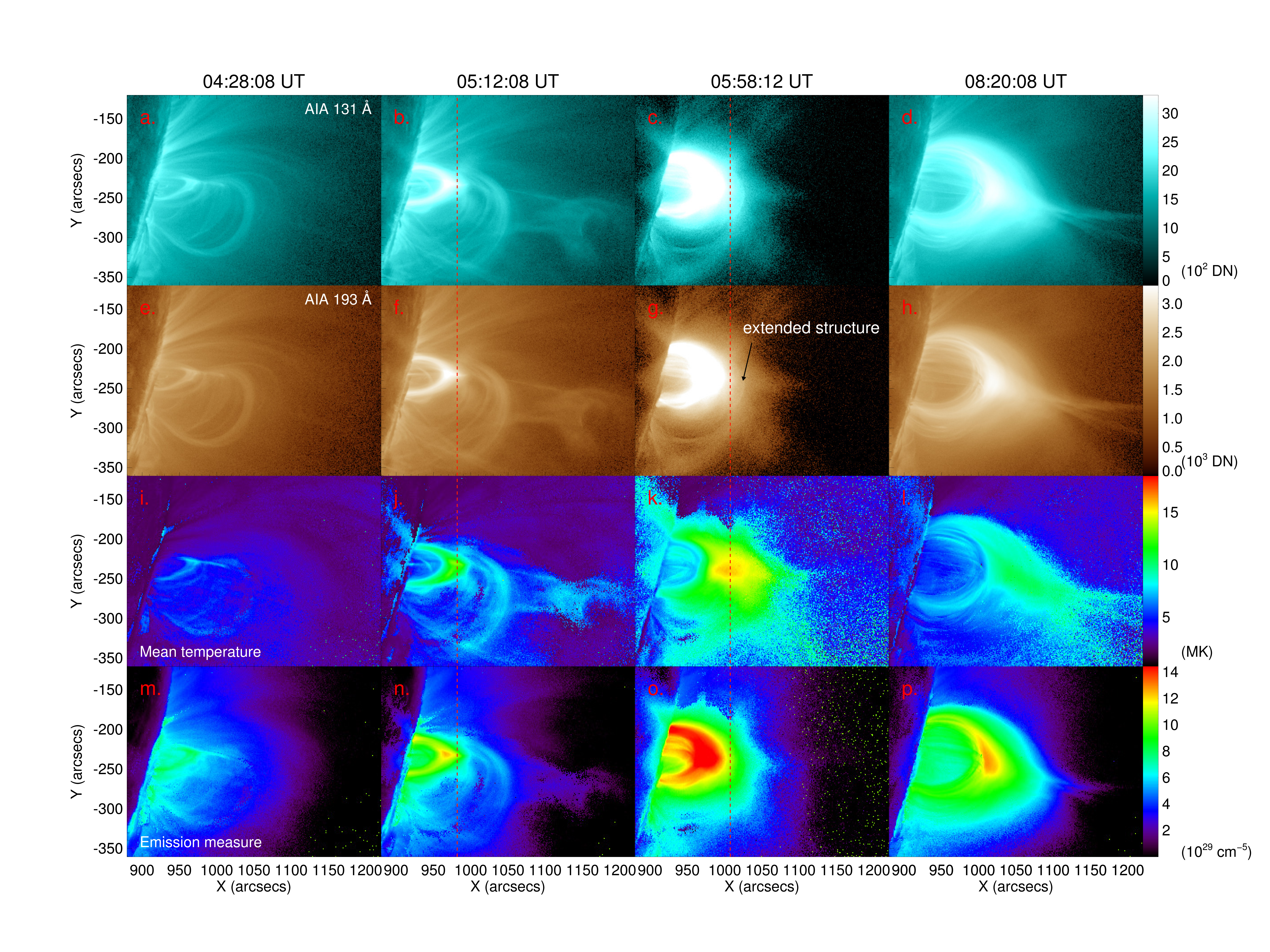}
               }
\caption{AIA 131 {\AA} images (a)--(d), 193 {\AA} images (e)--(h), temperature maps (i)--(l), and EM maps (m)--(p) for the flare region. The vertical dash lines are used to compare the height of different structures. Values of the maximum temperature, mean temperature, maximum EM, and mean EM are displayed in the corresponding images.} \label{f2}
(Animations of this figure are available in the online journal.)
\end{figure*}

\begin{figure*} 
     \vspace{-0.0\textwidth}    
     \centerline{\hspace*{0.00\textwidth}
               \includegraphics[width=0.8\textwidth,clip=]{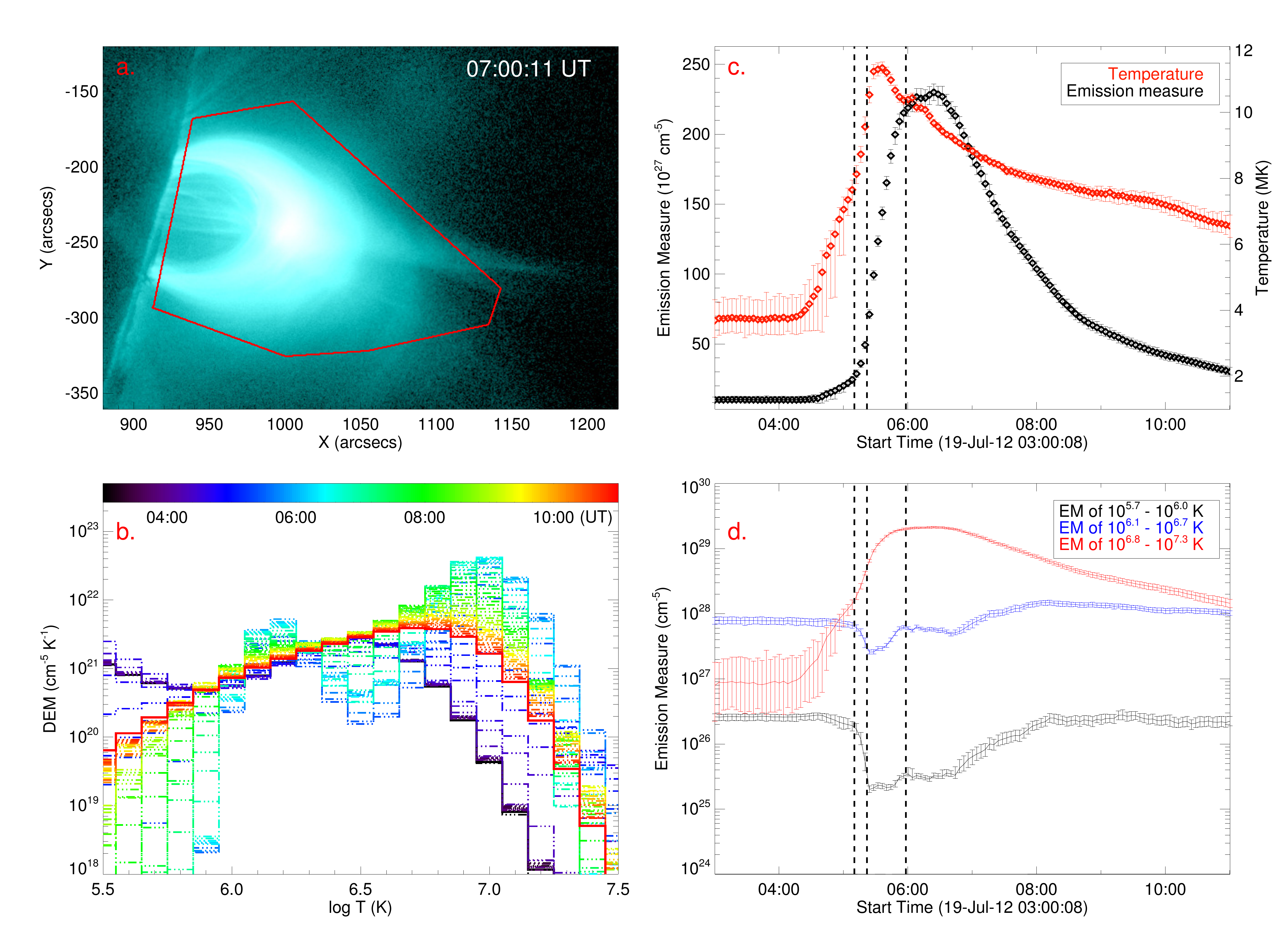}
               }
\caption{ DEM of the whole flare region. (a) AIA 131 {\AA} image at 07:00:11 UT with the red frame showing the area for integration. (b) The DEM curve varying with time, (c) time evolutions of the EM and mean temperature, and (d) the EMs integrated at three temperature intervals: 5.7$\leq$ log${T}\leq$ 6.0, 6.1$\leq$ log${T}\leq$ 6.7, and 6.8$\leq$ log${T}\leq$ 7.3, for the selected area. The error bars shown in (c) and (d) are determined from 100 Monte Carlo simulations.} \label{f3}
\end{figure*}

\begin{figure*} 
     \vspace{-0.0\textwidth}    
     \centerline{\hspace*{0.00\textwidth}
               \includegraphics[width=0.6\textwidth,clip=]{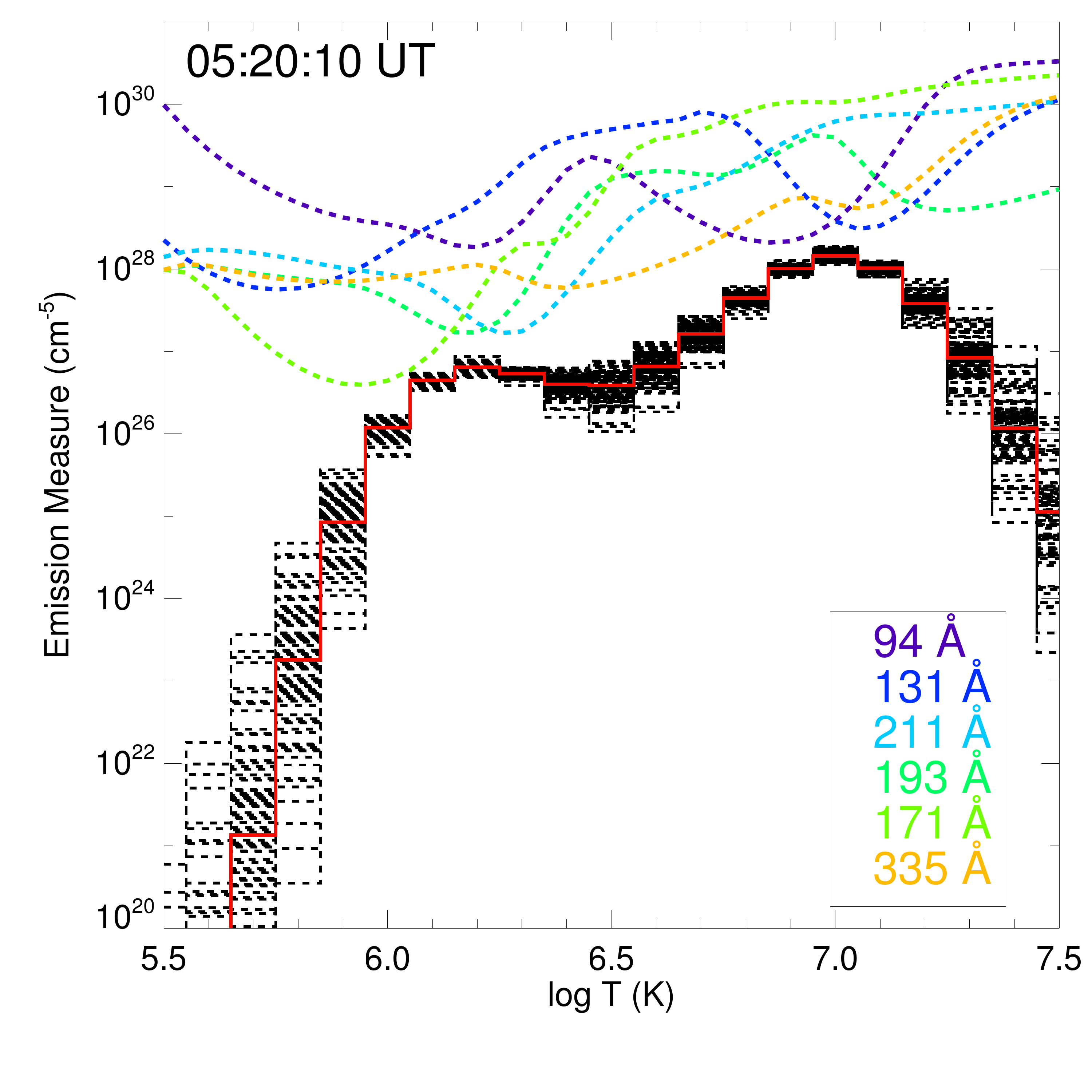}
               }
\caption{ Loci curves of the whole flare region at 05:20:10 UT. The red curves and the dash black curves are the optimal EMD and the EMDs from 100 Monte Carlo simulations, respectively. Here, the EM is calculated as ${\rm EM}={\rm DEM}\Delta T$, where $\Delta T=T\ln 10 \Delta(\log T)$ and $\Delta(\log T)$ is set to be 0.1.} \label{f4}
\end{figure*}

\begin{figure*} 
     \vspace{-0.0\textwidth}    
     \centerline{\hspace*{0.00\textwidth}
               \includegraphics[width=1.0\textwidth,clip=]{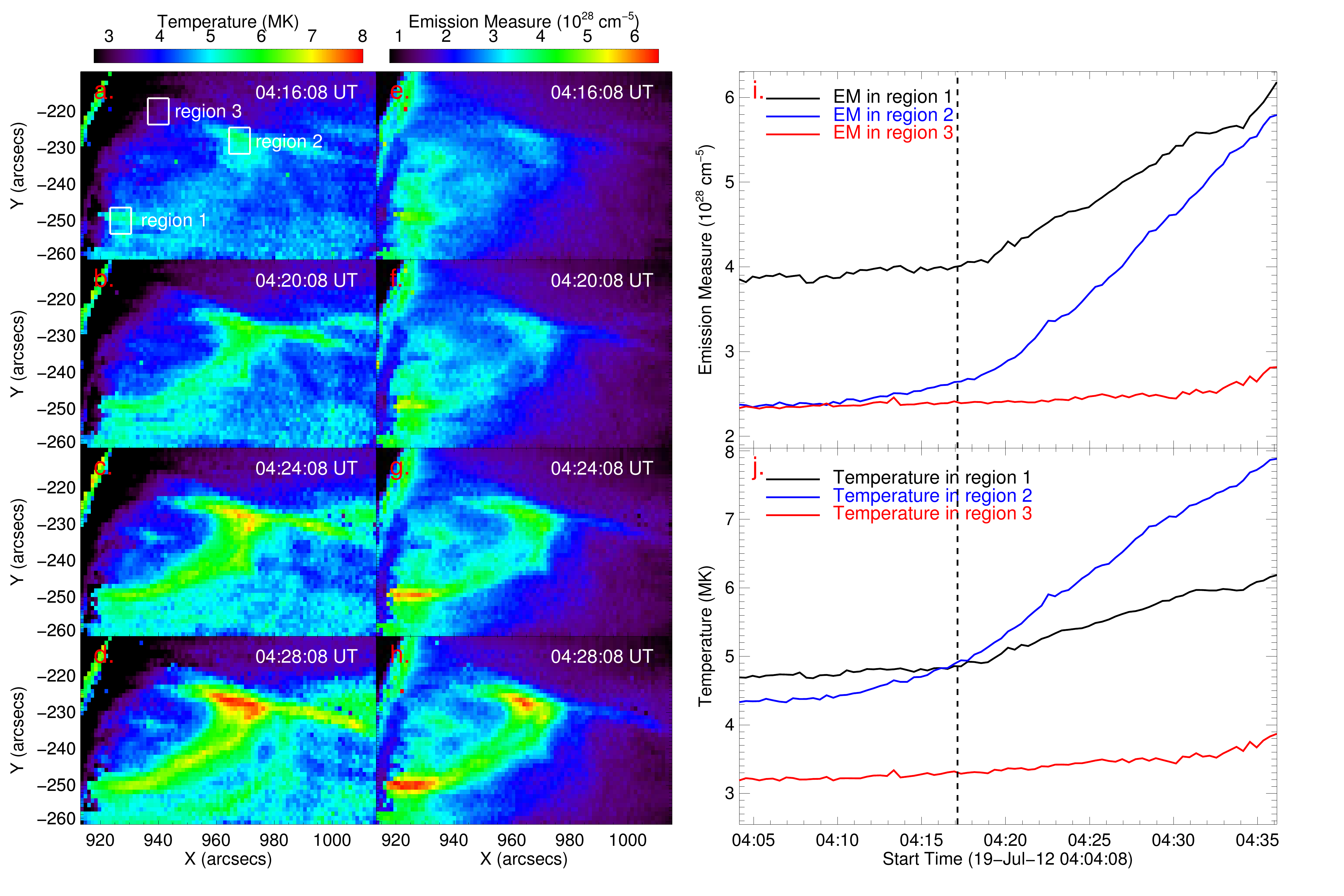}
               }
\caption{The DEM-weighted temperature maps (a)--(d) and EM maps (e)--(h) at four selected times. The three small boxes in (a) show the selected regions for evaluating the different EMs and temperatures. Evolutions of the EM (i) and DEM-weighted temperature (j) for the three selected regions.} \label{f5}
\end{figure*}

\begin{figure*} 
     \vspace{-0.0\textwidth}    
     \centerline{\hspace*{0.00\textwidth}
               \includegraphics[width=1.1\textwidth,clip=]{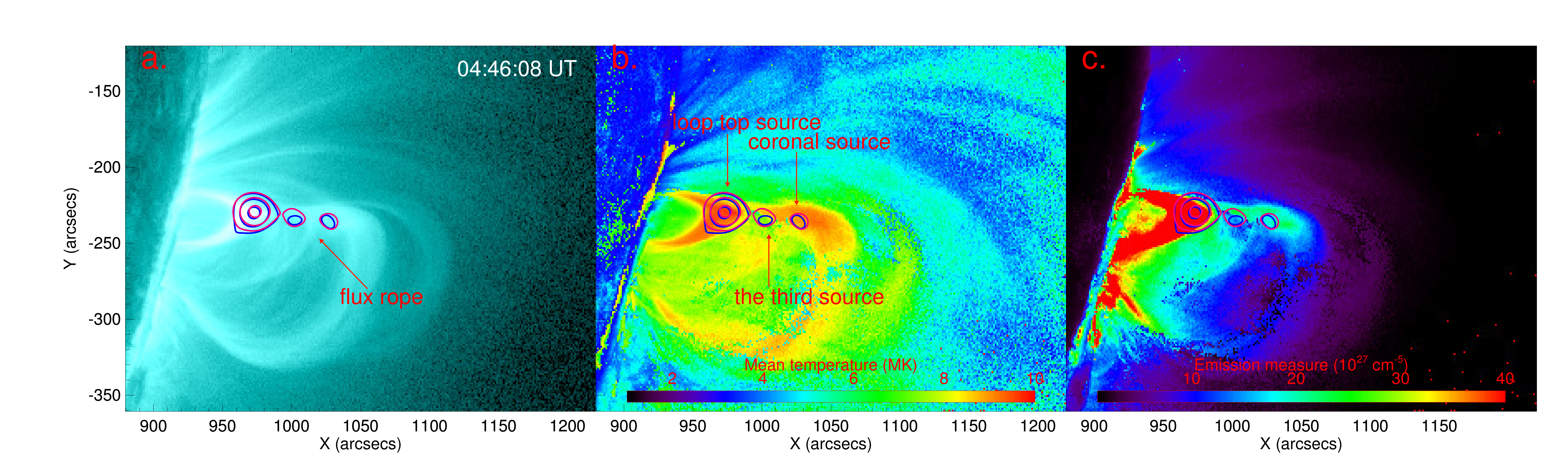}
               }
\caption{AIA 131 {\AA} image (a), temperature distribution (b), and EM distribution (c) of the magnetic reconnection region at about 04:46 UT. Over-plotted are the contours of RHESSI HXR emission. The contour levels are 30\%, 55\%, and 90\% for 6--10 keV, and 33\%, 55\%, and 90\% for 10--15 keV, respectively.} \label{f6}
\end{figure*}

\begin{figure*} 
     \vspace{-0.0\textwidth}    
     \centerline{\hspace*{0.00\textwidth}
               \includegraphics[width=1.1\textwidth,clip=]{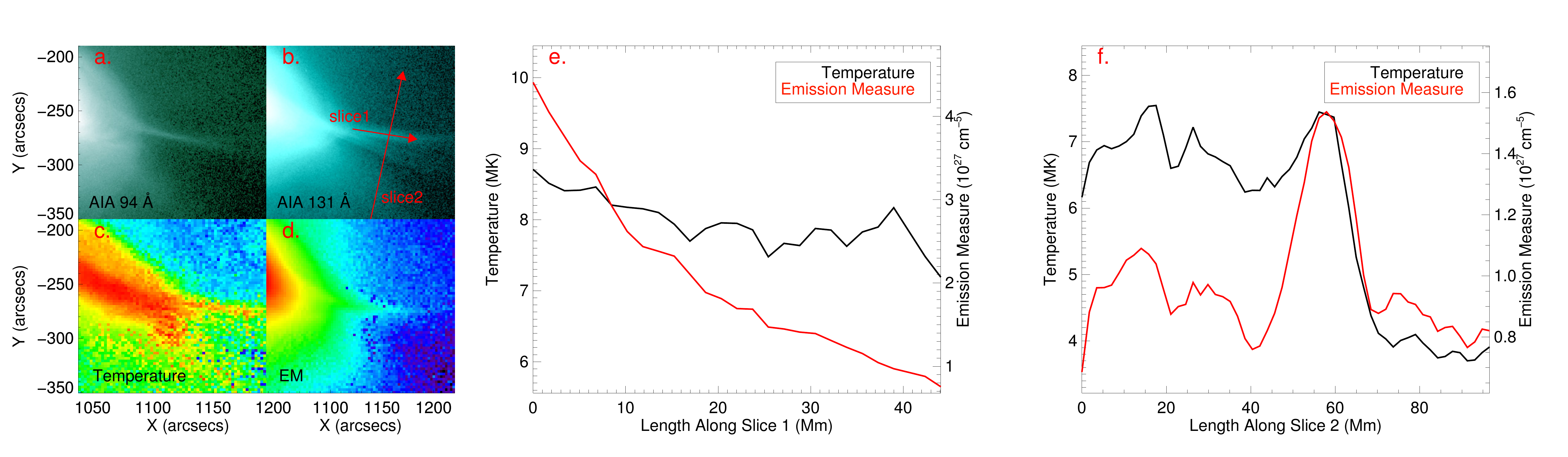}
               }
\caption{(a)--(d) Images of 94 {\AA}, 131 {\AA}, temperature, and EM at about 08:40 UT. The two slices in (b) indicate two directions along and perpendicular to the possible current sheet. (e) The temperature and EM along slice 1. (f) The temperature and EM along slice 2.} \label{f7}
\end{figure*}

\begin{figure*} 
     \vspace{-0.0\textwidth}    
     \centerline{\hspace*{0.00\textwidth}
               \includegraphics[width=0.9\textwidth,clip=]{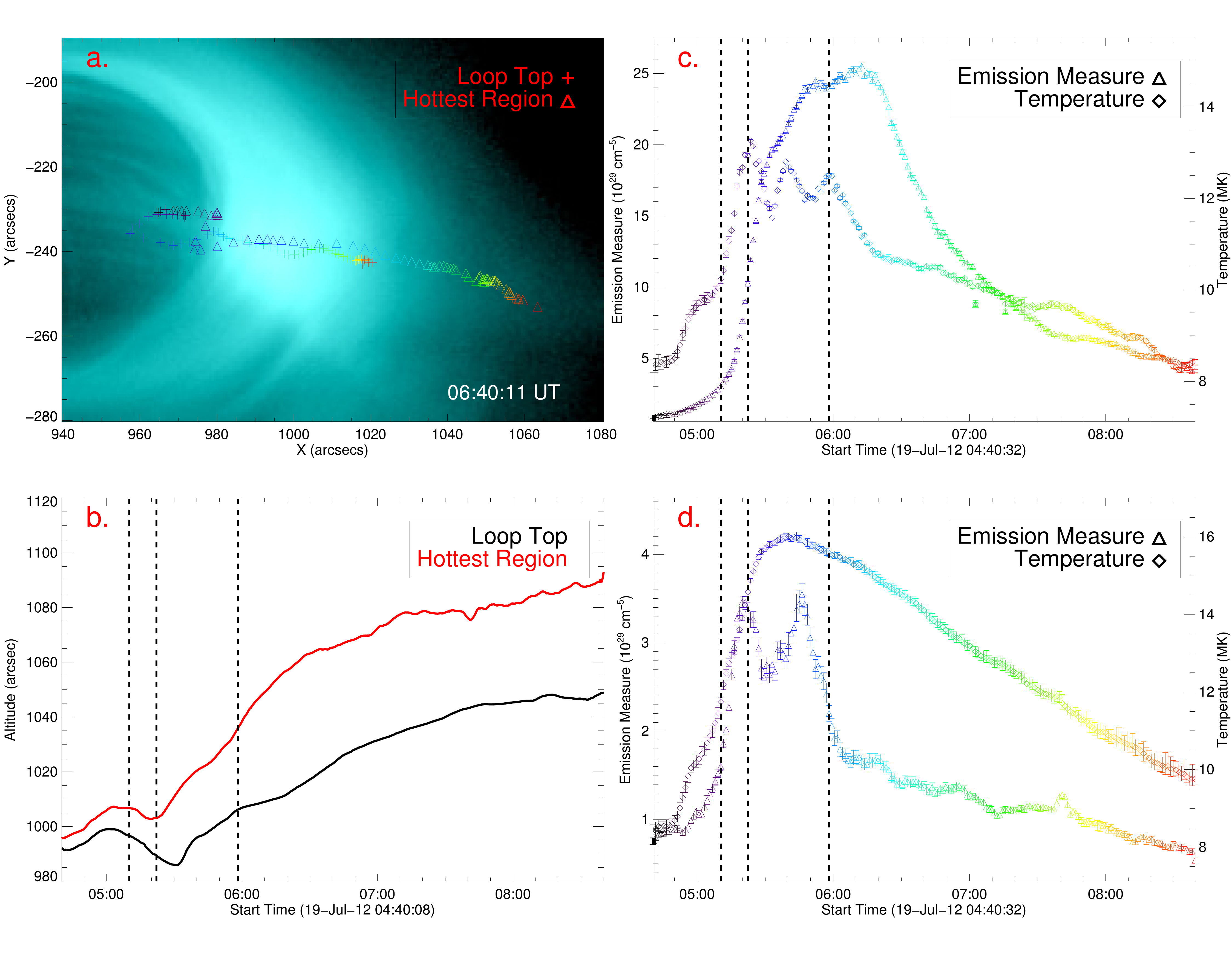}
               }
\caption{(a) AIA 131 {\AA} images at 06:40:11 UT, with the loop top and the hottest region trajectories over-plotted as colored crosses and triangles, respectively. (b) Temporal variations of the loop top and the hottest region altitude.  (c) Temporal evolutions of the temperature and EM in the loop top. (d) Temporal evolutions of the temperature and EM in the hottest region.  The error bars shown in (c) and (d) are determined from 100 Monte Carlo simulations.} \label{f8}
\end{figure*}

\begin{figure*} 
     \vspace{-0.0\textwidth}    
     \centerline{\hspace*{0.00\textwidth}
               \includegraphics[width=0.85\textwidth,clip=]{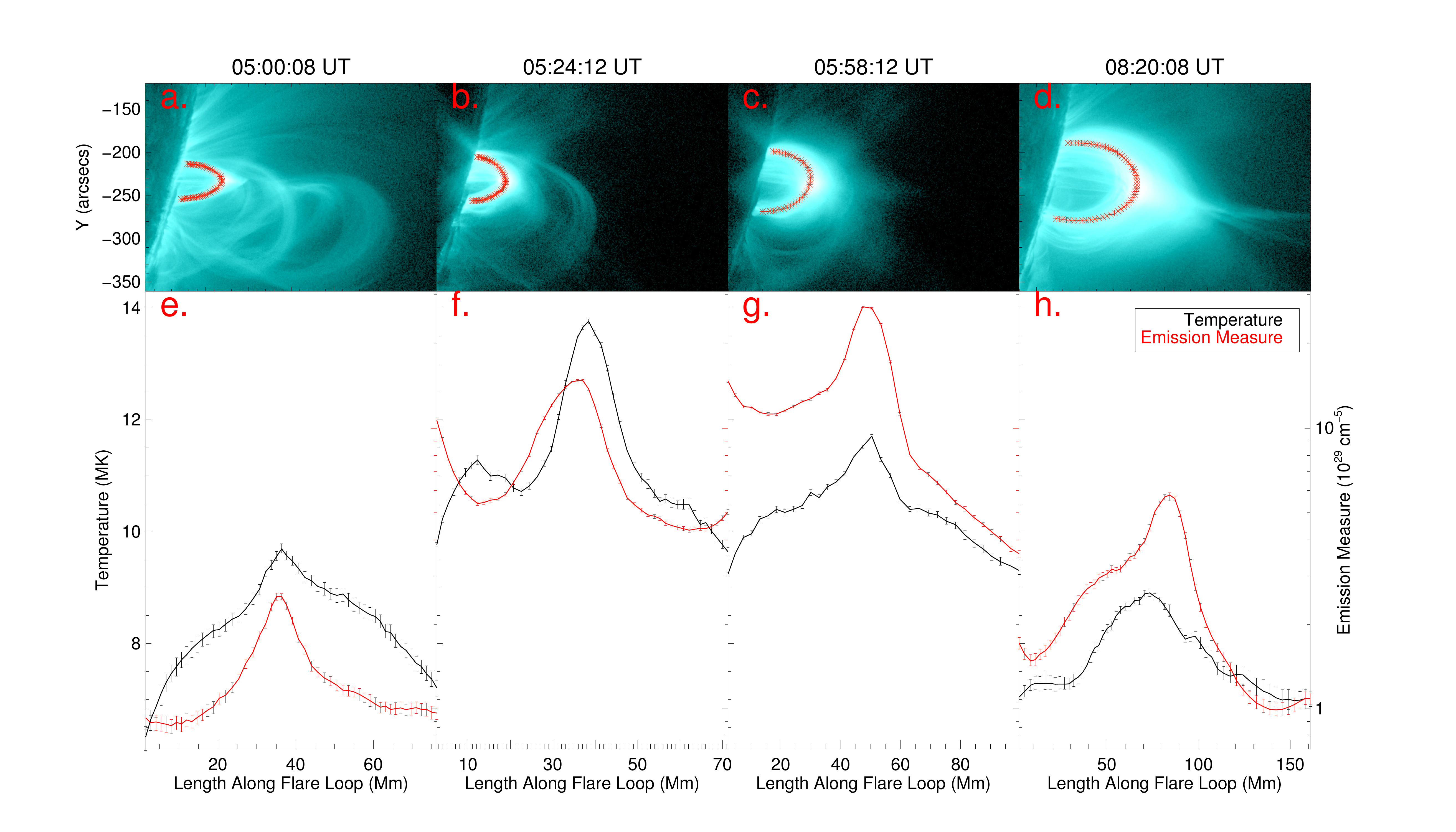}
               }
\caption{(a)--(d) AIA 131 {\AA} images at four selected times. The red stars are selected points along the flare loop for DEM computations. (e)--(h) The temperature and EM distributions of the selected points. The length is measured from the northern footpoint to the southern footpoint.} \label{f9}
\end{figure*}

\begin{figure*} 
     \vspace{-0.0\textwidth}    
     \centerline{\hspace*{0.00\textwidth}
               \includegraphics[width=0.8\textwidth,clip=]{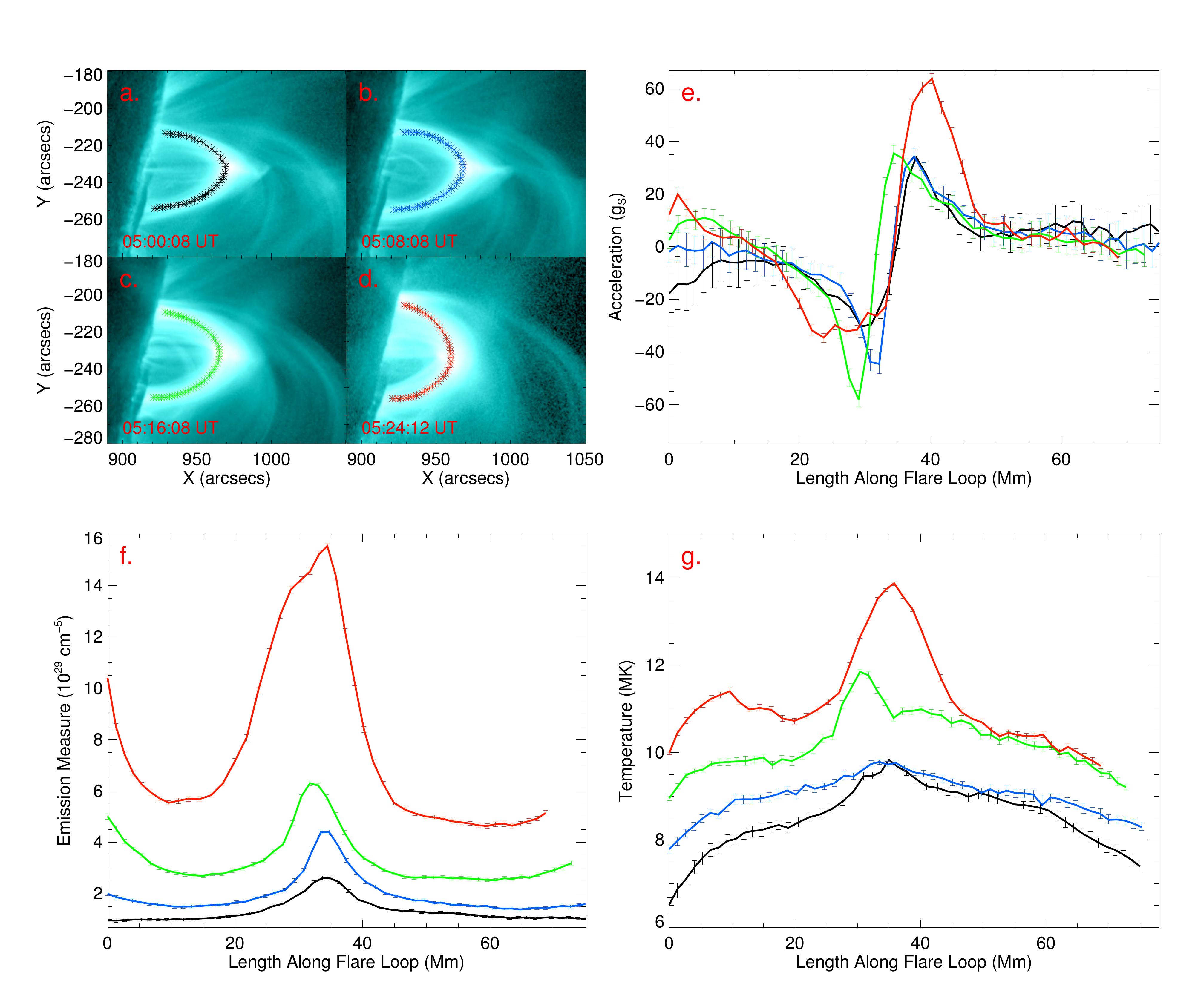}
               }
\caption{(a)--(d) AIA 131 {\AA} images at 05:00:08 UT, 05:08:08 UT, 05:16:08 UT, and 05:24:12 UT. The colored stars denote the points along the flare loop for computations. Distributions of the plasma acceleration in units of the gravitational acceleration at the solar surface (e), EM (f), and temperature (g) along the flare loops at the four selected times. The error bars are determined by making 100 Monte Carlo simulations.} \label{f10}
\end{figure*}

\end{document}